\documentclass{emulateapj}
\usepackage{natbib}
\usepackage[usenames,dvipsnames]{color}
\usepackage{amsmath}

\shorttitle{WD COMPANIONS OF BLUE STRAGGLERS}
\shortauthors{GOSNELL ET AL.}

\begin{document}
\title{Detection of white dwarf companions to blue stragglers in the open cluster NGC 188: direct evidence for recent mass transfer}

\thanks{This is WIYN Open Cluster Survey paper LVIII}

\author{Natalie M. Gosnell\altaffilmark{1}, Robert D. Mathieu\altaffilmark{1}, Aaron M. Geller\altaffilmark{2,3}, Alison Sills\altaffilmark{4}, Nathan Leigh\altaffilmark{5,6}, Christian Knigge\altaffilmark{7}}
\email{gosnell@astro.wisc.edu}
\altaffiltext{1}{Department of Astronomy, University of Wisconsin - Madison, 475 N. Charter St., Madison, WI  53706}
\altaffiltext{2}{Center for Interdisciplinary Exploration and Research in Astrophysics (CIERA) and Department of Physics and Astronomy, Northwestern University, 2145 Sheridan Rd, Evanston, IL  60208}
\altaffiltext{3}{Department of Astronomy and Astrophysics, University of Chicago, 5640 S. Ellis Avenue, Chicago, IL 60637}
\altaffiltext{4}{Department of Physics and Astronomy, McMaster University, 1280 Main St. W, Hamilton, ON  L8S 4M1, Canada}
\altaffiltext{5}{Department of Physics, University of Alberta, CCIS 4-183, Edmonton, AB  T6G 2E1, Canada}
\altaffiltext{6}{Department of Astrophysics, American Museum of Natural History, Central Park West and 79th Street, New York, NY 10024}
\altaffiltext{7}{School of Physics and Astronomy, University of Southampton, Highfield, Southampton, SO17 IBJ, UK}

\begin{abstract}
Several possible formation pathways for blue straggler stars have been developed recently, but no one pathway has yet been observationally confirmed for a specific blue straggler. Here we report the first findings from a \textit{Hubble Space Telescope} ACS/SBC far-UV photometric program to search for white dwarf companions to blue straggler stars. We find three hot and young white dwarf companions to blue straggler stars in the 7-Gyr open cluster NGC 188, indicating that mass transfer in these systems ended less than 300 Myr ago.  These companions are direct and secure observational evidence that these blue straggler stars were formed through mass transfer in binary stars.  Their existence in a well-studied cluster environment allows for observational constraints of both the current binary system and the progenitor binary system, mapping the entire mass transfer history. 
\end{abstract}


\section{INTRODUCTION}
Blue straggler stars (BSSs), originally defined to be anomalous stars more luminous and bluer than the main sequence (MS) of a star cluster, are now known to be present in open star clusters \citep{Johnson55}, globular clusters \citep{Sandage53,Ferraro99}, the Galactic field \citep{Preston00}, and dwarf spheroidal galaxies \citep{Momany07}.  They define an alternative stellar evolution path, yet not a rare one; BSSs comprise 25\% of the evolved population in NGC 188 \citep{Geller08}. Even so, the formation of BSSs has been a puzzle for almost six decades.  Recent work has identified several possible formation pathways, including mergers in hierarchical triples \citep{Perets09}, collisions during dynamical encounters \citep{Knigge09, Leigh11}, and mass transfer on the red giant branch (RGB) or asymptotic giant branch \citep[AGB;][]{Chen08}.  Yet no one pathway has yet been observationally confirmed for a specific BSS.  

The open cluster NGC 188 contains one of the most thoroughly studied populations of BSSs in our Galaxy.  Spectroscopic studies reveal that 80\% of the BSSs in NGC 188 are in binaries, and their remarkable period and eccentricity distributions provide clues to their formation histories \citep{Mathieu09,Mathieu13}. Of the 16 NGC 188 BSSs in binaries, all but three have orbital periods near 1000 days.  Statistical analysis of their secondary-mass distribution suggests that the companions to these BSSs have masses around 0.5 M$_{\odot}$ \citep{Geller11}.  Such masses suggest white dwarf (WD) companions, whose presence would indicate mass transfer as the dominant BSS formation mechanism in NGC 188. 

The presence of a WD is detectable as an excess of UV emission above the expected emission from a BSS alone.  In this Letter we present the first results from a \textit{Hubble Space Telescope} (\textit{HST}) program searching for WD companions to the BSSs in NGC 188.  We outline the observational design in \S~\ref{obs}, the WD detections in \S~\ref{detections}, implications of these detections in \S~\ref{discussion}, mass transfer histories in \S~\ref{masstransfer}, and summarize our results in \S~\ref{summary}.

\section{OBSERVATIONS}
\label{obs}
NGC 188 contains 20 BSSs\footnote{After the \textit{HST} observations we discovered a $V$-band photometric error for WOCS 1947, previously categorized as a BSS, which revealed it to be a RGB star.  It is no longer considered a member of the NGC 188 BSS population \citep{Mathieu13}.} determined to be three-dimensional members from radial-velocity (RV) and proper-motion (PM) studies \citep{Platais03,Geller09}.  The population consists of 2 double-lined spectroscopic binaries, 14 single-lined spectroscopic binaries, and 4 stars that do not show velocity variations.  In addition, there is one BSS candidate (WOCS 4230) that is a PM member and velocity variable but whose rapid rotation does not permit sufficiently precise velocities for a secure RV membership determination. We observed the 14 single-lined binaries, the 4 non-velocity variable BSSs, and WOCS 4230 using the Advanced Camera for Surveys (ACS) Solar Blind Channel (SBC) onboard \textit{HST}.  The non-velocity variable BSSs may still have very wide binary companions beyond the RV sensitivity, so they are included in our observational study.  We did not observe the two double-lined BSS binaries as they have almost equal-mass companions that are already known to not be WDs.  Each BSSs was observed in a separate visit of 2 orbits, for a total of 38 orbits (GO:12492, PI: Mathieu), using the F140LP, F150LP, and F165LP filters with total exposure times of 2040 s, 2380 s, and 1564 s, respectively.  

\subsection{Aperture Photometry}
\label{phot}
We use the astrodrizzle routine to create master drizzled images in each filter for each target with a resulting pixel scale of 0.025\arcsec\ pixel$^{-1}$.  We carry out aperture photometry using the IRAF package \textsc{daophot} with an aperture radius of 6 pixels, or an angular radius of 0.15\arcsec. We then apply an encircled energy fraction correction to our count rates based on aperture photometry of normalized TinyTim-modeled point sources\footnote{\texttt{http://www.stsci.edu/hst/observatory/focus/TinyTim}}.  We apply the same drizzling routine to the TinyTim source and carry out aperture photometry, from which we calculate the fraction of flux missed by our aperture.  We find and apply encircled energy corrections of 0.83, 0.84, and 0.85 for F140LP, F150LP, and F165LP, respectively.  We calculate our corrected count rates by dividing the measured count rate by the encircled energy correction. We convert our corrected count rates to instrument fluxes using the conversion factors given in the ACS Data Handbook \citep{acshandbook}.

The nested bandpasses of the \textit{HST}/ACS/SBC long-pass filters provide an opportunity to isolate the bluest UV flux.  As shown in previous studies \citep[e.g.,][]{Dieball05}, one can difference the count rates from the long-pass filters to create derived narrow bandpasses.  We create the derived filters of F140N = (F140LP$-$F150LP) and F150N = (F150LP$-$F165LP).  To photometer the count rates for these derived filters we difference the corrected long-pass filter count rates and calculate fluxes by applying conversion rates of
{\small
\begin{align*}
\mathrm{PHOTLAM_{SBC/F140N}}&=7.0631\times10^{-17}\: \mathrm{erg\: cm^{2}\: \AA^{-1}\: count^{-1}} \\
\mathrm{PHOTLAM_{SBC/F150N}}&=6.4868\times10^{-17}\: \mathrm{erg\: cm^{2}\: \AA^{-1}\: count^{-1}}.
\end{align*}
}%
 
We find the conversion rates using the IRAF package \textsc{synphot} and the tasks \textsc{calcband} and \textsc{bandpar}.  After finding the narrow-band fluxes we calculate magnitudes in all filters using 
$\mathrm{STMAG} = -2.5\,\mathrm{log}_{10}(\mathrm{flux})-21.1.$

\section{WHITE DWARF DETECTIONS}
\label{detections}
\begin{figure}
\begin{center}
\includegraphics[scale=1.0]{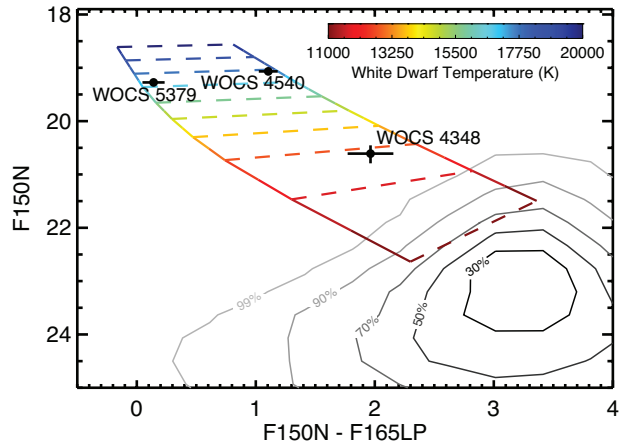}
\end{center}
\caption{Far-UV color-magnitude diagram showing the positions of WOCS 4540, 5379, and 4348, each shown with 2-sigma error bars. The grey contours show the expected density distribution of single BSSs with temperature and luminosity distributions matching the BSS population in NGC 188 \citep{Mathieu09}.  The outermost contour encompasses 99\% of single BSSs. The colored lines track synthetic BSS-WD pairs with increasing WD temperature as indicated by the color bar.  The lower and upper solid lines follow binaries with BSS temperatures of 6,000 K and 6,500 K, respectively \citep{uvblue}. The dashed lines track constant WD temperature from 11,000 K to 20,000 K in increments of 1,000 K (P. Bergeron, private communication). Comparing the observations with the synthetic BSS-WD models we determine that all three systems have WD companions with temperatures of approximately 13,000 K and above, which corresponds to an age of less than 300 Myr.}
\label{uvcmd}
\end{figure}

We detect a high-significance UV excess from WDs associated with three of the BSS: WOCS 4348, 4540, and 5379.  (We will present the full BSS population in a future paper.)  These binaries have periods of $1168\pm8$ days, $3030\pm70$ days, and $120.21\pm0.04$ days, respectively. In Figure~\ref{uvcmd} we show the far-UV color-magnitude diagram (CMD), utilizing the derived narrow bandpasses described in Section~\ref{phot}. The data are shown as black dots with 2-sigma error bars.  The grey contours show the expected CMD distribution for single BSSs.  We apply a reddening of E($B-V$)=0.09 \citep{Sarajedini99} to spectral models that match the temperature and luminosity range of BSSs in NGC 188 \citep{uvblue,Geller09}.  The resulting range of models is Monte Carlo-sampled 10,000 times and synthetically observed using the \textsc{synphot} package to create the density contours plotted in Figure~\ref{uvcmd}.  The outermost contour contains 99\% of the expected single-BSS distribution.  The three detected BSSs are much brighter and bluer than the single-BSS distribution; they must have an additional source of UV emission.  

\begin{figure*}
\begin{center}
\includegraphics[scale=0.75]{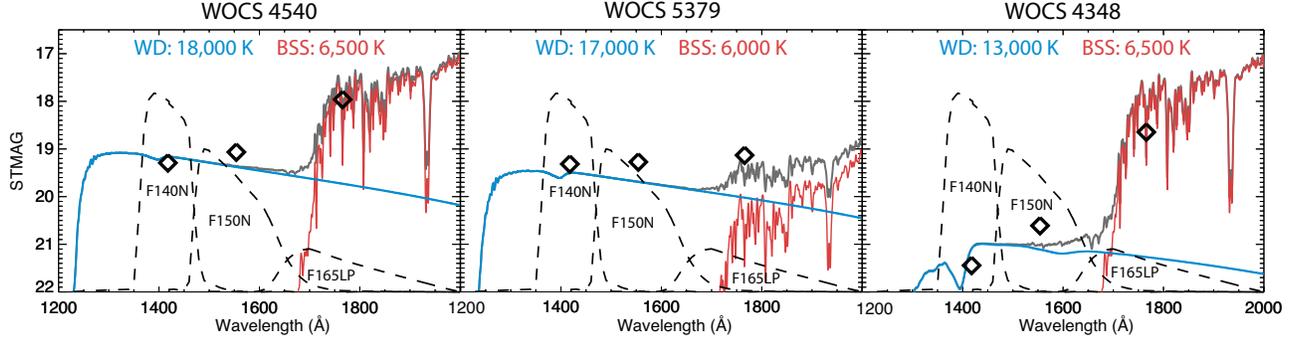}
\end{center}
\caption{Model spectra for the BSS-WD binaries compared to the observed narrow-band photometry, shown with black diamonds. The 2-sigma errors on the data are approximately the size of the symbols.  In each case the BSS emission \citep[modeled as a MS atmosphere,][]{uvblue} is shown in red, the WD emission (P. Bergeron, private communication) is shown in blue, and the combined spectrum is shown in gray.  The temperature components are indicated on the figure, and are the closest fit to the data given the precision in our temperature grid.  The dashed lines show the derived narrow-band filters used in this study.  It is apparent from this figure that the UV excesses detected for the BSS binaries are well explained by a hot WD companion.}
\label{uvspec}
\end{figure*}

To illustrate that the UV fluxes for these sources are straightforwardly explained by the presence of young, hot WD companions, in Figure~\ref{uvspec} we show model spectra of a BSS-WD binary compared to the narrow-band photometry for each of the three sources. The WD and BSS components are shown in blue and red, respectively, with the combined spectrum shown in grey.  The photometric data are shown as black diamonds.  The BSS temperatures shown are consistent with temperatures found from broadband fits to each BSS spectral energy distribution. 

\section{DISCUSSION}
\label{discussion}

\subsection{WD Companions as Evidence of BSS Mass Transfer Formation}

We argue that these young WDs are the result of very recent mass transfer that consequently created the BSSs. Binaries with periods from approximately 100 days to 1000 days are the expected result of mass transfer begun when the primary star has evolved onto the RGB or AGB and expanded in radius \citep[][\S~\ref{masstransfer}]{Chen08}. The primary star overflows its Roche lobe and transfers its envelope onto a MS companion, creating a BSS. Alternatively, mass can be accreted from stellar winds, especially during the AGB phase of evolution. In both cases, the core of the evolving primary remains at the end of the transfer of the envelope, which we now observe as a WD. Thus these three binaries are the first direct observational determination of the formation mechanism for specific BSSs. Furthermore, the age of the WD dates the formation time of the associated BSS.

WDs cool as they age.  Thus, the UV color can be used as both a temperature and an age diagnostic.  The colored lines in Figure~\ref{uvcmd} track synthetic observations of BSS-WD binaries with increasing WD temperature (with WD $\log_{10}(g/{\rm cm\,s^{-2}}) = 7.75$; P. Bergeron, private communication).  Comparing these tracks to our observations shows that all three WD companions are hotter than 12,000 K.  WD cooling models \citep{Holberg06,Tremblay11} indicate that a temperature of 12,000 K corresponds to a maximum cooling age of about 300 Myr for these WDs. 

\begin{figure*}
\begin{center}
\includegraphics[scale=0.8]{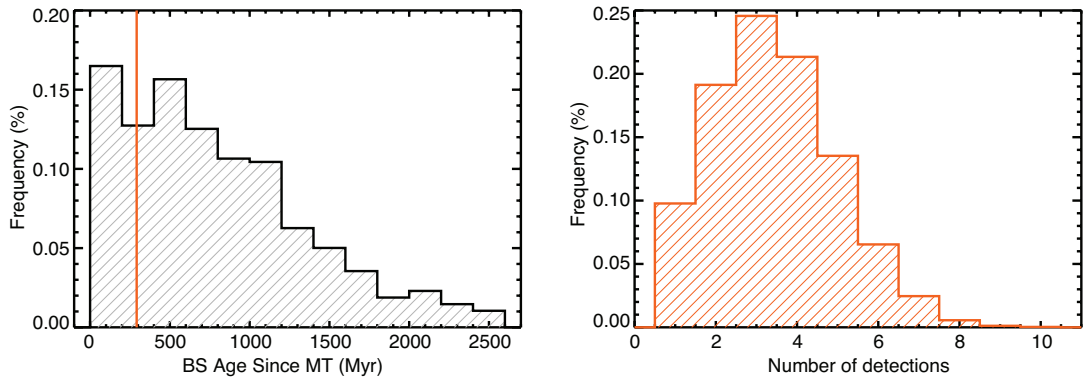}
\end{center}
\caption{Histogram of ages for mass transfer-formed BSSs in a sophisticated $N$-body model of NGC 188 \citep[left;][]{Geller13}, and a histogram of WD detection frequencies given our observational design and the $N$-body age distribution (right). The vertical line corresponds to the age of a 12,000 K WD, which is the detection limit given our observational design.  The histogram on the right is a result of Monte Carlo-sampling the age distribution. We expect $3.4\pm1.5$ detections, which is consistent with the 3 detections presented in this Letter.}
\label{agedist}
\end{figure*}

The detection of three hot WD companions is consistent with all 14 single-lined BSS binaries having formed via mass transfer.  We find an age distribution for mass transfer-formed BSSs from an $N$-body model of NGC 188 \citep{Geller13}. The age distribution is created by tracking the age since the end of mass transfer (which we take to be the WD age) for all mass transfer-formed BSSs in 20 realizations of NGC 188. The \textit{HST} observations are able to detect WD companions with temperatures down to 12,000 K, corresponding to an age of 300 Myr \citep{Holberg06,Tremblay11}.  We calculate the expected number of detections by Monte Carlo-sampling the age distribution and determining how many of the BSSs in a 14-binary sample would have WD companions hot enough to detect.  We expect $3.4\pm1.5$ detections, which is consistent with the 3 detections presented in this Letter (see Figure~\ref{agedist}.)  If the 4 non-velocity-variable BSSs are in wide binaries (increasing the sample to 18) we would expect $4.2\pm1.8$ detections, still consistent with these results.

\subsection{Dynamical Confusion?}
We rule out the possibility of dynamical encounters altering our observed binaries since their formation. For present-day cluster parameters we adopt: a core radius of 1.3 pc; a one-dimensional velocity dispersion of 0.41 km s$^{-1}$; a total of 1470 cluster members in NGC 188 (objects down to $V=21$ with PM membership probability $\ge10$\%); and a giant star population that comprises 8\% of the total cluster \citep{Platais03,Geller09}. We take the current giant population as a liberal proxy for all WDs with ages of less than 300 Myr. To convert the core luminosity density to a mass density we use a mass-to-light ratio of 1.5 \citep{deGrijs08}. We use binary fractions from an $N$-body model of NGC 188 that does not suffer from observational incompleteness \citep{Geller13}. Using the formulae for calculating dynamical encounters found in \citet{Leigh11}, we find it would take $9.2\times10^{9}$ years for any BSS binary to have an exchange encounter with another single giant star, and it would take $4.1\times10^{10}$ years for any BSS binary to have an encounter with another binary containing a giant.  Both of these values exceed the age of NGC 188, and well exceed the maximum 300 Myr age of our BSS-WD binaries. We do not expect any of the WD companions discovered here to be the result of a dynamical exchange. 

We also rule out the possibility that the BSS-WD binaries have gone through \textit{any} dynamical interaction, which would cause the observed orbital period to not equal the period at the end of mass transfer. It would take $1.4\times10^{9}$ years for any BSS-WD binary to interact with any single star.  It would take $5.6\times10^{8}$ years for any BSS-WD binary to interact with any average binary system \citep[period of 2200 days;][]{Geller13} in the cluster.  As these time scales also exceed the 300 Myr age of our systems we do not expect the BSS-WD binaries to have undergone dynamical interactions of any kind since their formation.

\section{MASS TRANSFER HISTORIES}
\label{masstransfer}
The confluence of very recent mass transfer, well-defined binary parameters, and a thoroughly studied open cluster provides a unique opportunity to define the formation paths of these BSS-WD systems in some detail, including constraining their progenitor binaries. Here we lay out the remarkable amount of empirical knowledge known for these systems and provide preliminary results for formation paths.

Specifically, for these BSS-WD binaries we know: 1) the mass transfer ended in the last 300 Myr; 2) the mass transfer occurred during the RGB or AGB phases of the primary star evolution, because the current companions are WDs; 3) the progenitor primary star had a MS mass of 1.155--1.185 M$_{\odot}$, the mass range before wind loss of NGC 188 RGB and AGB stars over the past 300 Myr \citep{Sarajedini99}; 4) the progenitor secondary star had a mass greater than approximately 0.8 M$_{\odot}$, so that after mass transfer (for common assumptions about giant mass loss and non-conservative mass transfer) the BSS mass is greater than 1.1 M$_{\odot}$, the current MS turnoff mass; and 5) the orbital period at the end of mass transfer is the current orbital period.

Additionally, the final orbital period after Roche lobe overflow (RLOF) is related to the giant core mass at the end of mass transfer \citep{Rappaport95}.  This is due to the monotonic relationship between giant radius and core mass at the point when the envelope detaches from the core \citep{Joss87,Smedley14}.  The giant core mass is the WD mass in the final binary system. Broadly, the WD will be more massive for longer-period final binaries since the progenitor primary must evolve further to have the larger radius necessary to undergo RLOF and will therefore have a more massive core.

We use this wealth of information to qualitatively discuss the formation paths of each of the BSS-WD binaries, informed by a preliminary exploration of the progenitor-binary parameter space using the Binary-Stellar Evolution code \citep[BSE;][]{Hurley02}. A complete theoretical analysis is beyond the scope of this Letter, but as a proof-of-concept we present here one specific BSE formation path for each binary.  We provide three digit precision for the masses to allow for independent modeling by interested readers, but given the parameterized nature of BSE we note that more sophisticated models of these binaries will certainly modify the values given below. 

\subsection{Specific Formation Scenarios}

\textit{WOCS 4348}:  With a final orbital period of $1168\pm8$ days, mass transfer via RLOF occurs while the progenitor primary star is on the AGB. BSE can produce WOCS 4348 starting with a binary having a 1.182 M$_{\odot}$ primary and a 1.118 M$_{\odot}$ secondary with an orbital period of 1684.87 days.  When mass transfer begins the primary has been reduced to 0.9 M$_{\odot}$ due to substantial mass loss from stellar winds.  After non-conservative mass transfer, the final binary has a Carbon/Oxygen (CO) WD of mass 0.574 M$_{\odot}$ and a BSS of mass 1.288 M$_{\odot}$ with an orbital period of 1168.25 days.  The mass transfer ends at an age of 6.754 Gyr resulting in a WD age of 246 Myr (temperature of 12,712 K). 

\textit{WOCS 5379}: With an orbital period of $120.21\pm0.04$ days, the progenitor primary star begins RLOF before it leaves the RGB. BSE produces WOCS 5379 with a progenitor binary comprised of a 1.167 M$_{\odot}$ primary and a 1.122 M$_{\odot}$ secondary with an orbital period of 468.77 days.  Mass transfer commences on the RGB, and completes at an age of 6.932 Gyr with a Helium WD of mass 0.439 M$_{\odot}$ and a BSS of mass 1.135 M$_{\odot}$.  The final binary period is 120.32 days with a WD age of 68 Myr (temperature of 18,236 K).  This BSE scenario does go through a short phase of common envelope (CE) evolution.  There is evidence indicating that the BSE parameterized treatment of CE may create more CE systems than seen in observations \citep{Geller13}, and so we caution that this scenario may be BSE-dependent.

\textit{WOCS 4540}: With a period of $3030\pm70$ days, the longest of all the BSSs in NGC 188, 
the final binary is difficult to produce by traditional RLOF \citep{Chen08}, but can be created through wind accretion.  In this scenario, WOCS 4540 will have a CO WD companion with a mass of about 0.55 M$_{\odot}$ based on single-star evolution theory. BSE begins with a 1.174 M$_{\odot}$ primary and a 1.061 M$_{\odot}$ secondary with an orbital period of 2564.93 days. Wind accretion begins in earnest as the primary ascends the AGB, but the primary star radius never exceeds the Roche lobe radius.  The primary core is exposed at 6.923 Gyr, for a CO WD age of 77 Myr (temperature of 17,630 K). The original secondary star accretes enough mass to reach 1.162 M$_{\odot}$ and is observed as a BSS with a 0.539 M$_{\odot}$ WD companion.  The final orbital period is 3029.84 days. 

These three BSSs span the range of $V$-band brightness of BSSs in NGC 188. WOCS 4348 lies 0.3 mags above the MS turnoff, while WOCS 5379 is fainter but bluer than the turnoff \citep{Mathieu09}, both reasonable for the BSE-suggested masses of 1.288 M$_{\odot}$ and 1.135 M$_{\odot}$, respectively. However, WOCS 4540 is one of the most luminous blue-straggler binaries in NGC 188, unexpected for the BSE-suggested mass of only 1.162 M$_{\odot}$. This may cast doubt on the wind mass-transfer scenario, or perhaps the AGB wind prescription in BSE needs modification.  The dense, slow wind produced throughout the normal AGB evolution of the primary star may result in more efficient mass accretion than is implemented in BSE \citep{Abate13}. However, the presence of very luminous BSSs has always been a challenge for mass-transfer models \citep{Chen08} and indeed have been cited as arguments for alternative merger and collision theories. The presence of the very young WD companion to WOCS 4540 revitalizes the long-standing question regarding the BSS mass-luminosity relation \citep[e.g.;][]{Sandquist03,Geller12}.  

\section{SUMMARY}
\label{summary}
Using \textit{HST} far-UV photometry we detect three young WD companions to BSSs in NGC 188.  These BSS binaries formed through mass transfer within the past 300 Myr.  This is the first direct determination of the formation mechanism for a specific population of BSSs.  The three detections presented here are consistent with most or all the NGC 188 single-lined BSS in binaries being formed through mass transfer.  The existence of these systems within a well studied cluster environment allows for observational constraints of the progenitor binary parameters, setting the timeline for the entire mass transfer process.  These binaries will serve as important test cases to constrain more sophisticated mass transfer modeling efforts in the future.

\acknowledgements We thank our referee for the insightful comments.  We also acknowledge Pierre Bergeron for providing the WD model spectra. N.~M.~G. and R.~D.~M. are supported through HST Program number 12492, provided by NASA through a grant from the Space Telescope Science Institute, which is operated by the Association of Universities for Research in Astronomy, Incorporated, under NASA contract NAS5-26555.  A.~S. is supported by the Natural Sciences and Engineering Research Council of Canada. A.~M.~G. is funded by a National Science Foundation Astronomy and Astrophysics Postdoctoral Fellowship under Award No. AST-1302765. 

{\it Facilities:} \facility{HST (ACS)}
\\
\\

\end{document}